# Quasi-Large Hole Polarons in BiVO₄: Implications for Photocatalysis and Solar Energy Conversion

Zhimeng Hao, Taifeng Liu*

National & Local Joint Engineering Research Center for Applied Technology of Hybrid Nanomaterials, Henan University, Kaifeng 475004, China

Corresponding author: Taifeng Liu: tfliu@vip.henu.edu.cn

Abstract: BiVO₄ is a promising photocatalyst for solar energy conversion, but its efficiency is limited by small polaron formation. However, some physical properties of BiVO₄ deviate from typical small polaron behavior. Using the state-of-the-art first-principles calculations, we demonstrate that BiVO₄ forms a quasi-large hole polaron with a radius around 2 nm, resembling free carriers with high mobility. This polaron is stabilized primarily by acoustic phonon modes, creating a shallow trap state near the valence band maximum, which prolongs its lifetime. Simultaneously, it retains a redox potential comparable to that of free carriers. We propose that such large polarons explain the superior properties of BiVO₄ and other transition metal oxide photocatalysts. Tuning phonon modes to stabilize large polarons offers a promising strategy for designing materials with enhanced solar energy conversion efficiency.

**TOC Graphic**

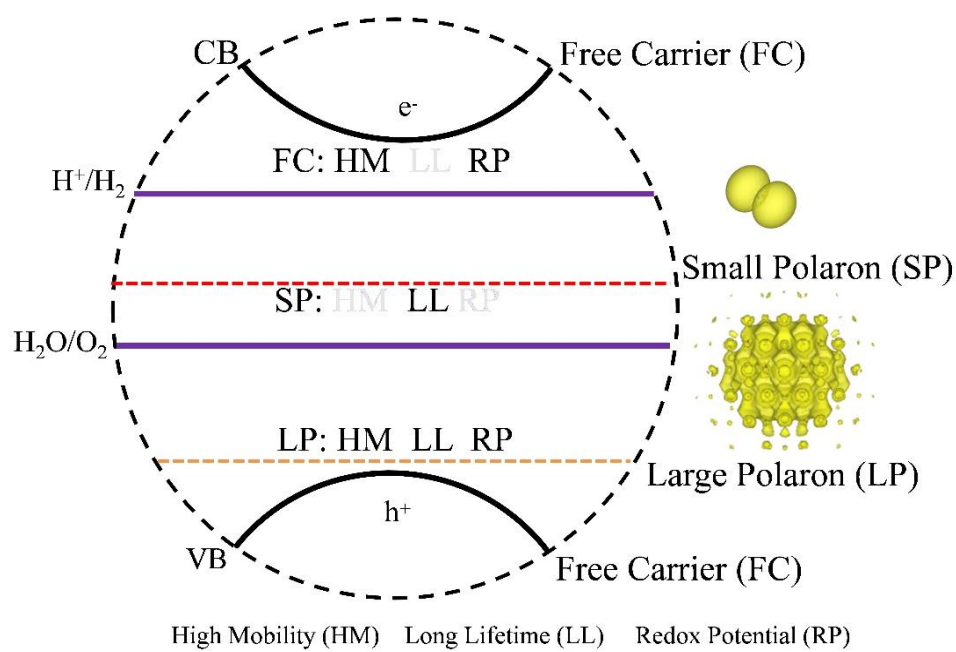

Bismuth vanadate (BiVO$_4$) has gained significant attention as a photocatalyst and photoelectrochemical material for water splitting, enabling efficient solar-to-chemical energy conversion[1-5]. Its advantages include abundant and inexpensive elements[6], a suitable bandgap for light absorption[7], and favorable band edges for water oxidation[8-10]. Key factors affecting solar energy conversion efficiency—such as charge carrier mobility, lifetime, and diffusion length—determine charge separation and transport[11, 12]. Time-resolved microwave conductivity (TRMC) experiments report a carrier mobility of 0.04 cm$^2$ V$^{-1}$ s$^{-1}$, a lifetime of 40 ns, and a diffusion length of 70 nm in BiVO$_4$[13], contributing to its excellent solar energy conversion performance. These properties are strongly mediated by polaron formation, a phenomenon prevalent in BiVO$_4$ and other transition metal oxides[14].

Both electron and hole polarons in BiVO$_4$ have been extensively studied through theoretical and experimental approaches. For electrons, density functional theory (DFT) with on-site Coulomb corrections (DFT+U) and hybrid DFT methods consistently show electron localization on vanadium atoms, forming small polarons[15, 16]. This self-trapping induces significant local lattice distortions, reducing V$^{5+}$ to V$^{4+}$.[17] The small polaron exhibits an activation energy of ~0.35 eV for nearest-neighbor hopping, with mobility on the order of 10$^{-4}$ cm$^2$ V$^{-1}$ s$^{-1}$[18, 19]. This localization creates a deep trap state approximately 1 eV below the conduction band minimum (CBM).

For hole carriers in BiVO$_4$, localization differs across computational approaches. Using DFT+U, the hole self-traps on a single oxygen atom, forming a small hole polaron with significant lattice distortion and a charge transition from O$^{2-}$ to O$^{-}$. In contrast, hybrid functionals predict a hole localized on the BiO$_8$ dodecahedron, with ~20% of the charge on the Bi atom, characterized by Bi 6s orbitals[20, 21]. Both types are classified as small polarons, forming deep trap states (~1 eV above the valence band maximum, VBM) primarily from O 2p or hybridized Bi 6s orbitals[16, 22]. The small hole polaron has an activation energy of ~0.3 eV for hopping and a mobility on the order of 10$^{-4}$ cm$^2$ V$^{-1}$ s$^{-1}$[23]. Recent several studies have revealed that holes in BiVO$_4$ exhibit a more delocalized state compared to electrons. Sun et al.[24] employed pump-probe transient reflection microscopy to directly observe electron and hole dynamics in BiVO$_4$. Their findings reveal that electrons are more localized, while holes exhibit greater delocalization. Seo et al.[25] investigated the hole polaron in BiVO$_4$ using hybrid DFT with a 4×4×2 supercell (768 atoms). They revealed that the self-trapped energy of hole polaron is -0.1 eV which may result in free carriers in the valence band (VB).

Additionally, their study found that with smaller supercells (3×3×1 and 3×3×2), holes do not localize.

Current research suggests small polaron formation in BiVO$_4$. However, this model is inconsistent with experimental measurements[24], as small polaron hopping predicts much lower mobility. Furthermore, deep trap states associated with small polarons impair water oxidation performance. The conclusions of small polaron formation heavily depend on computational parameters such as the U value in DFT+U or the Hartree-Fock exchange fraction (α) in hybrid functionals.

In this study, we investigated polarons in BiVO$_4$ without employing the supercell approach, eliminating the need for DFT+U or hybrid functionals. we employ a state-of-the-art first-principles approach accounting for electron-phonon coupling to investigate polarons in BiVO$_4$. The ab initio polaron equations were solved to accurately model isolated polarons[16]. We identify a quasi-large hole polaron which could explain the experimental findings. This polaron forms a shallow trap state, enhancing its lifetime without compromising water oxidation capability.

The ground state wave function $\psi(r)$ and atomic displacements $\Delta\tau_{\kappa\alpha p}$ forming a polaron can be found by minimizing the total DFT energy functional of an excess electron added to a crystal, which translates into the solution of the following coupled system of equations[26],

$$\hat{H}^0_{KS}\psi(r) + \sum_{\kappa\alpha p} \frac{\partial V^0_{KS}}{\partial \tau_{\kappa\alpha p}} \Delta\tau_{\kappa\alpha p}\psi(r) = \varepsilon\psi(r) \quad (1)$$

$$\Delta\tau_{\kappa\alpha p} = -\sum_{\kappa'\alpha'p'} (C^0)^{-1}_{\kappa\alpha p, \kappa'\alpha'p'} \int dr \frac{\partial V^0_{KS}}{\partial \tau_{\kappa'\alpha'p'}} |\psi(r)|^2 \quad (2)$$

$\tau_{\kappa\alpha p}$ represents the Cartesian coordinate of the atom $k$ in the unit cell $p$ along the direction $\alpha$, $C^0_{\kappa\alpha p, \kappa'\alpha'p'}$ is the matrix of interatomic force constants[27], and $\hat{H}^0_{KS}$ and $V^0_{KS}$ represent the Kohn-Sham Hamiltonian and the self-consistent potential, respectively. The superscript ⁰ indicates that the quantities are evaluated in the ground state without extra electron. We will refer to $\varepsilon$ as the polaron eigenvalue[28].

We can transform Eqs. (1) and (2) into a coupled set of equations for the expansion coefficients in reciprocal space,

$$\frac{2}{N_p} \sum_{qmv} B_{qv} g^*_{mnv}(K,q) A_{mk+q} = (\varepsilon_{nk} - \varepsilon) A_{nk} \quad (3)$$

$$B_{qv} = \frac{1}{N_p} \sum_{mnk} A^*_{mk+q} \frac{g_{mnv}(k,q)}{\hbar \omega_{qv}} A_{nk} \quad (4)$$

In these expressions, $\varepsilon_{nk}$ are the Kohn-Sham eigenvalues, $\omega_{qv}$ are the phonon frequencies and $g_{mnv}(K,q)$ are the electron-phonon matrix elements[29, 30].

The polaron formation energy $\Delta E_f$, defined as the energy required to trap a conduction band state with eigenvalue $\varepsilon_{CBM}$ into a localized polaron. It consists of electron part and phonon part, can be obtained from the expansion coefficients that solve Eqs. (3) and (4) by[26]:

$$\Delta E_f = \frac{1}{N_p} \sum_{nK} |A_{nk}|^2 (\varepsilon_{nk} - \varepsilon_{CBM}) - \frac{1}{N_p} \sum_{qv} |B_{qv}|^2 \hbar \omega_{qv} \quad (5)$$

$\varepsilon_{CBM}$ are the kohn-Sham eigenvalue of the conduction band bottom, $\hbar$ are the reduced Plank constant. $\varepsilon_{nk}$ are the Kohn-Sham eigenvalues, $\omega_{qv}$ are the phonon frequencies. $A_{nk}$ is the coefficient of the single-particle Kohn-Sham state used to expand the polaron wave function, and $B_{qv}$ is the coefficient of phonon eigenmodes with frequencies $\omega_{qv}$ to expand the atomic displacement[26].

We performed density functional theory (DFT) calculations using the Quantum ESPRESSO package[31-33] along with Wannier90[33-35] and EPW codes[29, 30, 36]. Calculations were based on the Perdew-Burke-Ernzerh (PBE) parameterization of the exchange-correlation potential within the generalized gradient approximation (GGA)[32, 33], using optimized norm-conserving Vanderbilt (ONCV) pseudopotentials and plane waves with a kinetic energy cutoff of 90 Ry[37-39]. Phonon frequencies and electron-phonon matrix elements were determined through density functional perturbation theory. Ground state electron and lattice dynamics were calculated using $4\times4\times8$ k-point and $2\times2\times4$ q-point grids, respectively. The electron energies, phonon frequencies, and electron-phonon matrix elements interpolated onto dense grids through Wannier-Fourier interpolation with $8\times8\times8$ k-points and $8\times8\times8$ q-points. To obtain maximally localized Wannier functions, we included a subset comprising the 10 lowest conduction bands investigating electron polarons, 2 and 24 highest valence bands investigate hole polarons on bismuth and oxygen sites, respectively. We calculate the formation of polarons using an 8×8×8 primitive cell containing 6144 atoms to ensure that the periodic replicas of the polaron are sufficiently separated.

We systematically investigated all types of polarons, including hole and electron polarons, in tetragonal BiVO₄ using first-principles calculations. It is modeled using a 12-atom primitive cell for both geometry optimization and phonon calculations. The calculated lattice parameters, a =b = c=6.93 Å, α=β=136.19⁰, γ=63.69⁰, agree well with experimental values[40]. For a hole polaron, localization can occur either at oxygen or bismuth sites, and both scenarios were investigated in our study. We first investigated the hole localized on the oxygen sites. The isosurface plot of the polaron wavefunction for a hole localized on oxygen sites is shown in Figure 1(a). Unlike DFT+U or hybrid functional results[16], which predict localization on a single oxygen site, our calculations reveal the hole is distributed mainly across eight oxygen sites with O 2p character. The polaron radius is around 2 nm, larger than the lattice parameters of a BiVO₄ unit cell. The hole polaron has a formation energy of -22 meV, indicating stability, and forms a shallow trap state 0.18 eV above the valence band maximum (VBM).

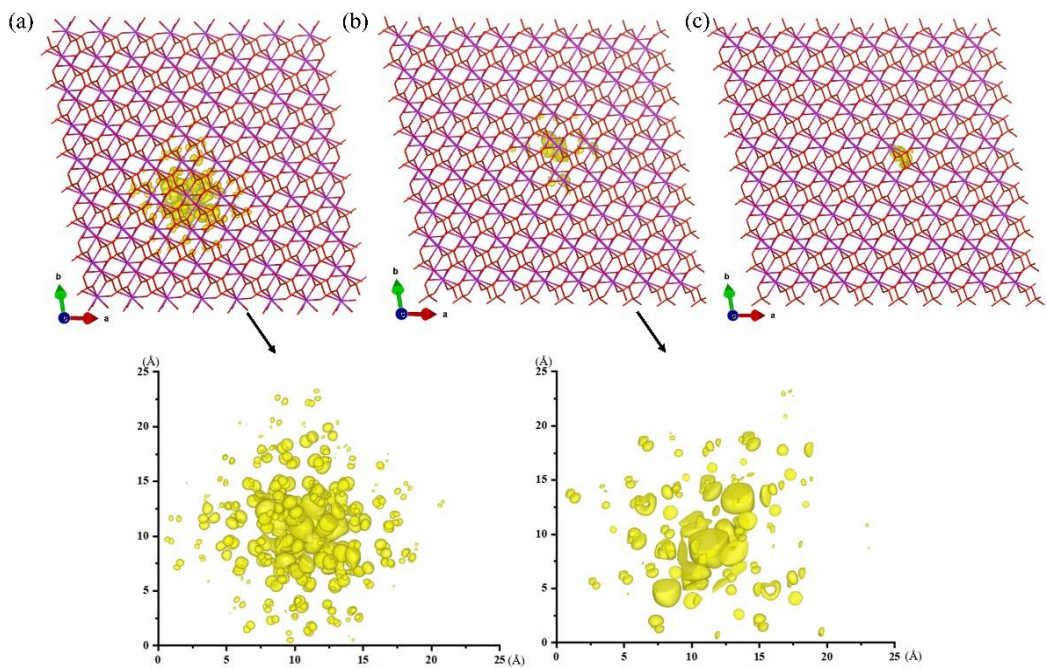

Figure 1 The isosurface plot of the hole polaron wavefunction on (a) O and (b) Bi sites with the insert figures to enlarge the polaron shape, and (c) the isosurface plot of the electron polaron wavefunction on V site

For the hole polaron on bismuth, the hole is primarily localized on two Bi atoms, with the remainder distributed across surrounding oxygen atoms, as shown in the isosurface plot of the polaron wavefunction in Figure 1(b). On Bi atoms, the hole occupies Bi 6s states, while on oxygen atoms, it resides in O 2p states. This result differs from hybrid functional calculations[41], which predict localization on a BiO$_8$ dodecahedron. Our findings indicate a more delocalized hole polaron, with a radius exceeding 2 nm, including the size of the contribution on oxygen sites. The formation energy of this polaron is -62 meV, making it more stable than the hole polaron on oxygen sites. The trap state is 0.32 eV above the valence band maximum (VBM), slightly deeper than that of the oxygen-site hole polaron.

We identified two types of hole polarons in BiVO$_4$, both with radii exceeding the unit cell lattice parameters. Compared to large polarons in materials like LiF[42] or halide perovskites[43], which have radius larger than 3 nm, the hole polaron in BiVO$_4$ is classified as a quasi-large polaron. With a radius between that of small and large polarons, its transport can involve hopping or free-carrier behavior, yielding mobility exceeding 1 cm²V⁻¹s⁻¹.

For electron carriers, a small polaron forms, localized on a vanadium site as shown in Figure 1(c). This result aligns with findings from both DFT+U and hybrid functional approaches, which consistently indicate that the electron forms a small polaron localized on the vanadium site. The radius of this small polaron is approximately 0.5 nm. This small polaron has a formation energy of -501 meV, significantly lower than that of the hole polaron, reflecting stronger localization. The trap state lies 1.43 eV below the CBM, deep within the 2.46 eV band gap, confirming its deep-trap nature. The activation energy for hopping is ~0.35 eV, with mobility in the 10⁻⁴ cm²V⁻¹s⁻¹ range[18, 19].

We investigated electron and hole polarons in BiVO$_4$ from the perspective of electron-phonon coupling. For hole polarons localized on oxygen sites, the electronic weights $|A_{nk}|^2$ superimposed on the band structure and the spectral function $A^2(E)$ are shown in Figure 2(a), while the corresponding atomic displacements resolved by $|B_{qv}|^2$ and the phonon spectral function $B^2(E)$ are shown in Figure 2(b). Similarly, the electron and phonon contributions for hole polarons localized on bismuth sites are displayed in Figures 2(c) and 2(d).

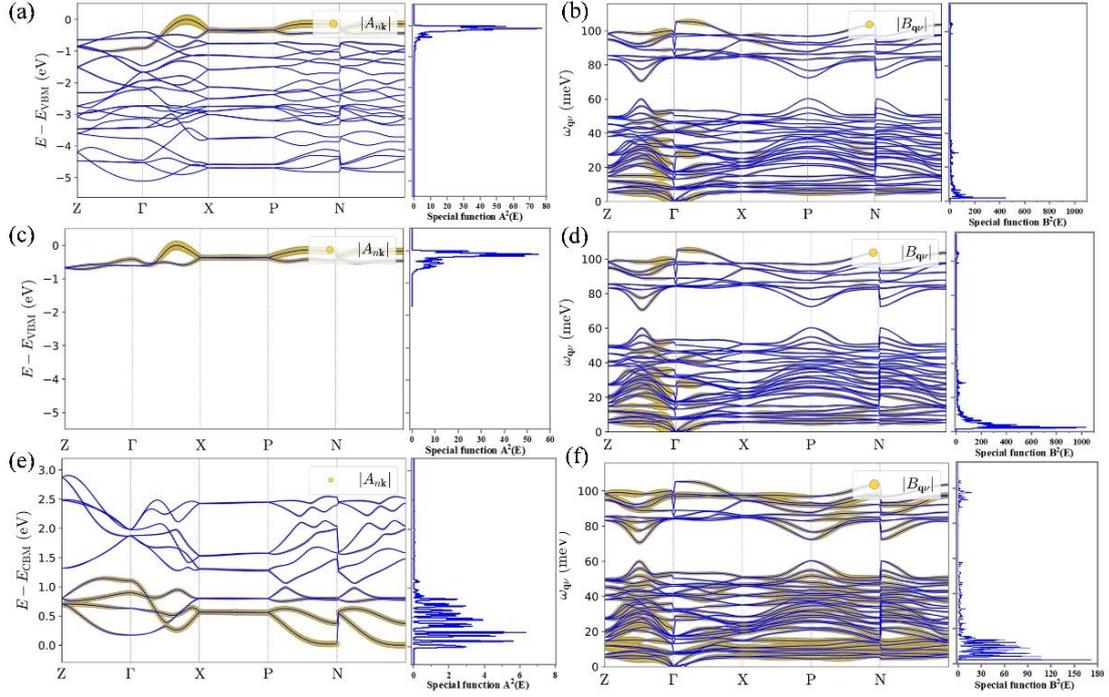

Figure 2 Generalized Fourier amplitudes $A_{nk}$ plotted on top of the phonon dispersion relations of BiVO₄ and spectral function $A^2(E)$. Generalized Fourier amplitudes $B_{qv}$ plotted on top of the phonon dispersion relations of BiVO₄ and spectral function $B^2(E)$. The hole polaron localization on (a) and (b) O sites, (c) and (d) Bi sites. (e) and (f) The electron polaron localization on V sites. The corresponding points of transverse wave vector k are Z(0.5, 0.5, -0.5), G(0.0, 0.0, 0.0), X(0.0, 0.0, 0.5), P(0.25, 0.25, 0.25), N(0.0, 0.5, 0.0).

Hole polarons are primarily associated with states at the top of the valence band, with minimal contributions from deeper valence band states. Phonon spectra reveal that transverse acoustic modes (<10 meV) dominate the coupling, with minor contributions from longitudinal optical (LO) modes at ~50 and ~100 meV. For oxygen-site hole polarons, phonon contributions amount to -0.16 eV, while electron contributions are 0.14 eV. For bismuth-site hole polarons, the respective contributions are -0.26 eV and 0.20 eV, explaining the greater stability of bismuth-site polarons. Notably, phonon energies forming bismuth-site polarons are nearly double those forming oxygen-site polarons.

For electron polarons localized on vanadium sites, Figures 2(e) and 2(f) show the electron and phonon contributions. The entire lowest conduction band contributes significantly to the polaronic wavefunction, with smaller contributions from higher bands. Coupling involves the entire LO phonon branch, with dominant contributions from modes below 20 meV and additional contributions from modes in the 80–100 meV range. The total phonon contribution is -0.87 eV, while the electron contribution is 0.42 eV, indicating strong electron-phonon coupling and small polaron formation for electrons. In contrast, the weaker electron-phonon coupling for holes leads to quasi-large polaron formation. These findings highlight distinct coupling mechanisms for electron and hole polarons in BiVO₄, with implications for charge transport and material performance.

Several experimental measurements are available for comparison with our calculated results. Early studies using TMRC[13] and THz [44] spectroscopy reported carrier mobilities on the order of $10^{-2}$ cm²V⁻¹s⁻¹, but these methods cannot distinguish between electrons and holes. Experimental mobility values are expected to be lower than theoretical predictions due to sample impurities and surface boundary effects, as our calculations assume a perfect crystal. Chi et al. [45] predicted the existence of large hole polarons in BiVO₄, attributed to O p orbitals hybridized with V d and Bi sp orbitals. Additionally, our calculated results demonstrate that holes form quasi-large polarons, directly supporting the findings of Sun et al.[24] and Seo et al.[25], which indicate that holes exhibit greater delocalization..

When the photocatalyst BiVO₄ absorbs photons, free carriers (electrons and holes) are generated, as depicted in Figure 3. These carriers exhibit high mobility and have conduction band (CB) and valence band (VB) levels suitable for redox reactions. However, free carriers typically have short lifetimes due to rapid recombination. Fortunately, they can transform into polarons, which act as trapped states with extended lifetimes.

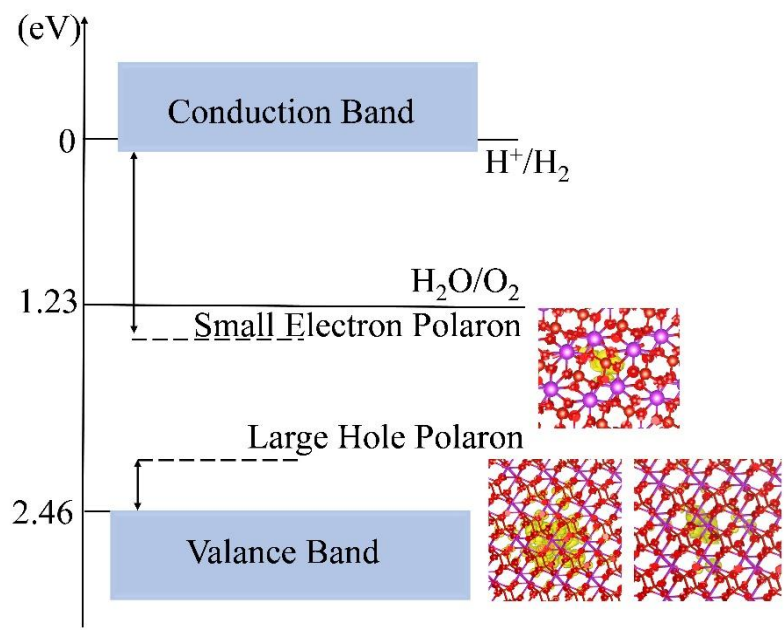

Figure 3 The energy level of the trap states of different types of polarons in the band gap of BiVO$_4$.

Small polarons are commonly observed in photocatalysts. While their lifetimes are longer, their mobility is extremely low ($10^{-4}$ cm$^2$V$^{-1}$s$^{-1}$). Additionally, small polarons often form deep trap states in the middle of the band gap, which can impair redox activity, as illustrated in Figure 3. Quasi-large or large polarons may also form in photocatalysts, although they remain less explored due to methodological limitations. Recent studies have reported large hole polarons in rutile TiO$_2$, quasi-2D electron polarons in anatase TiO$_2$[46], and potential large polarons in tungsten-based photocatalysts such as WO$_3$[47], Bi$_2$WO$_6$[48], and certain oxynitrides and nitrides[49, 50]. Large polarons exhibit mobility comparable to free carriers and form shallow trap states that minimally affect redox reactions. Most importantly, their trapped states have sufficiently long lifetimes to facilitate surface reactions.

In this study, we have identified a quasi-large hole polaron in BiVO$_4$. This hole polaron can be localized either on oxygen sites or on two bismuth sites along with the surrounding oxygen atoms. The radius of the polaron exceeds the lattice parameters of the BiVO$_4$ unit cell. It exhibits high mobility comparable to that of free carriers and forms a very shallow trap state close to the valence band maximum (VBM). The transverse acoustic phonon modes play a significant role in stabilizing this hole polaron. Based on these findings, we propose that quasi-large and large polarons are critical in photocatalysis, not only in BiVO$_4$ but also in other transition metal oxides. Their high mobility ensures efficient charge transport, while their shallow trap states provide extended lifetimes without compromising the redox capabilities of the material. Moreover, phonon modes significantly influence the formation and characteristics of large polarons. By controlling these phonon modes, it may be possible to tune the polaron radius, mobility, and trap states, paving the way for designing high-efficiency materials for solar energy conversion.

## ASSOCIATED CONTENT

**Notes**

The authors declare no competing financial interest.

**Data available**

The data that support the findings of this study are available from the corresponding author upon reasonable request.

ACKNOWLEDGMENT

This work was supported the Key Technologies R&D Program of Henan Province (No. 242102521002) and the National Natural Science Foundation of China (grant #22173026, 21703054).